\documentclass[twocolumn,trackchanges]{aastex62}

\shorttitle{SPOCS Kepler Sample}
\shortauthors{Brewer, Wang, Fischer, \& Foreman-Mackey}

\begin{document}

\title{Compact multi-planet systems are more common around metal poor hosts}

% =============================================================================
% Authors
%
\correspondingauthor{John M. Brewer}
\email{john.brewer@yale.edu}

\author[0000-0002-9873-1471]{John M. Brewer}
\affiliation{Department of Astronomy, Yale University, 52 Hillhouse Avenue, New Haven, CT 06511, USA}
\affiliation{Department of Astronomy, Columbia University, 550 West 120th Street, New York, New York 10027}
	\email{john.brewer@yale.edu}

\author[0000-0002-7846-6981]{Songhu Wang}
\affiliation{Department of Astronomy, Yale University, 52 Hillhouse Avenue, New Haven, CT 06511, USA}
	\email{songhu.wang@yale.edu}

\author[0000-0003-2221-0861]{Debra A. Fischer}
\affiliation{Department of Astronomy, Yale University, 52 Hillhouse Avenue, New Haven, CT 06511, USA}
	\email{debra.fischer@yale.edu}

\author[0000-0002-9328-5652]{Daniel Foreman-Mackey}
\affiliation{Flatiron Institute, 162 5th Ave, New York, New York 10010}
	\email{dforeman-mackey@flatironinstitute.org}

\begin{abstract}

In systems with detected planets, hot-Jupiters and compact systems of multiple planets are nearly mutually exclusive. We compare the relative occurrence of these two architectures as a fraction of detected planetary systems to determine the role that metallicity plays in planet formation. We show that compact multi-planet systems occur more frequently around stars of increasingly lower metallicities using spectroscopically derived abundances for more than 700 planet hosts.  At higher metallicities, compact multi-planet systems comprise a nearly constant fraction of the planet hosts despite the steep rise in the fraction of hosts containing hot and cool-Jupiters.  Since metal poor stars have been underrepresented in planet searches, this implies that the occurrence rate of compact multis is higher than previously reported. Due to observational limits, radial velocity planet searches have focused mainly on high-metallicity stars where they have a higher chance of finding giant planets.  New extreme-precision radial velocity instruments coming online that can detect these compact multi-planet systems can target lower metallicity stars to find them.

\end{abstract}

% =============================================================================
%  The Letter
%
\keywords{stars: solar-type}

% =============================================================================
%  Introduction
%
\section{Introduction} \label{sec:intro}

The gas giant planet-metallicity correlation \citep{2004A&A...415.1153S,2005ApJ...622.1102F} provided strong support for the core-accretion model \citep{1996Icar..124...62P} over gravitational instability: massive cores should form more rapidly in more metal rich (i.e. more massive) disks.  Because the correlation is for giant planets on short period orbits, radial velocity searches made it possible to develop clean samples of stars with and without high-mass planets.  As the number of discovered planets has exploded, we now know that almost all stars have planets \citep{2012ApJS..201...15H,2015ApJ...809....8B,2015ARA&A..53..409W}, making it impossible to know if a given star is planet-free or simply has planets that evade our limited detection capabilities.

The search for extrasolar planets has identified two notable system architectures in the region close to the host star: multiple small planets on tight orbits, compact multi-planet systems \citep{2011ApJS..197....8L} and massive planets on short orbits, hot-Jupiters.  These two system architectures are almost mutually exclusive, with few hot-Jupiters having close companions and almost no compact multi-planet systems having nearby massive planets.  Hot-Jupiters are uncommon, but occur more frequently around stars with high amounts of heavy elements \citep[high metallicity;][]{2004A&A...415.1153S,2005ApJ...622.1102F} but small planets can occur around stars with a wide range of metallicities \citep{2012Natur.486..375B,2014Natur.509..593B,2015AJ....149...14W,2018AJ....155...89P}.  

Compact multi-planet systems are mostly composed of planets near or below the current detection limits of radial velocity surveys.  However, they tend to be very co-planar \citep{2014ApJ...790..146F}, making them easy to detect in the \textit{Kepler} transiting planet survey \citep{2010Sci...327..977B}.  Some groups have looked at the average metallicity of these planets as a function of their radius and found that smaller planets are found around stars with a lower average metallicity \citep{2014Natur.509..593B,2015AJ....149...14W,2018MNRAS.480.2206O}.  This hints that protoplanetary disks with less available solids may struggle to form larger planets, or even planets at all, but issues of selection bias and detection completeness still obscure a complete picture.

To circumvent these problems, we chose to look only at systems with detected planets and compare the properties of systems with unique architectures. Planet detection should not be biased toward a given type of system at a particular metallicity. We derived uniform stellar properties and elemental abundances from high resolution spectra for almost 3000 stars, including almost 1200 planet hosts \citep{Brewer:2016gf,2018ApJS..237...38B} and compared the number of systems of a given architecture to all known hosts as a function of the host metallicity.  As stars evolve, the measured surface abundances of heavy elements can decrease \citep{2017ApJ...840...99D,2018ApJ...857...14S}, which might bias any analysis of the influence of those elements on planet architecture.  We limited our analysis to un-evolved stars ($\log g > 4.0$) and also looked at ratios of heavy elements to confirm our findings, since those ratios are relatively static over the main sequence lifetime of the stars.

% =============================================================================
%  Methods
%
\section{Data and Analysis}
\subsection{Stellar Properties and Abundances} \label{sec:abundances}
The stellar properties and abundances used for this Letter were all derived from high resolution optical spectra taken with the Keck HIRES spectrograph and analyzed in a homogeneous manner \citep{Brewer:2016gf,2018ApJS..237...38B}.  The analysis procedure has been shown to recover surface gravities consistent with those from asteroseismology to within 0.05 dex \citep{2015ApJ...805..126B} in addition to accurate temperatures and precise abundances for the abundances of 15 elements (C, N, O, Na, Mg, Al, Si, Ca, Ti, V, Cr, Mn, Fe, Ni, and Y).  Statistical uncertainties for the abundances range from $\sim 0.01$ dex (iron and silicon) to 0.04 dex (nitrogen), depending on the element.

The parameters and abundances are derived using forward modeling performed with the analysis package Spectroscopy Made Easy \citep[SME;][]{2017A&A...597A..16P}, fitting in an iterative fashion.  After continuum normalizing the spectrum and extracting 20 short wavelength segments totaling 350~\AA\ between 5160 and 7800~\AA, the initial temperature and gravity are set using broadband colors and the abundance pattern is set to that of the Sun. We then fit for the global stellar properties (effective temperature, surface gravity, rotational and Doppler broadening, metallicity) and the abundances of three $\alpha$ elements (Ca, Si, Ti) to allow for departures from the solar abundance pattern.  Using the derived parameters from this first fit, we perturb the temperature by $\pm 100$~K and re-fit, taking the $\chi^2$ weighted average as the global parameters from this stage.  The global properties are then fixed, and we solve for the abundances of the 15 elements.  This set of parameters and the abundance pattern are then used as a new starting point for a second iteration of the procedure.

The parameters and abundances are precise for dwarf stars of high signal-to-noise ratio (S/N), but trends in abundance with temperature have been identified for evolved stars and those with S/N $< 45$.  To avoid potential contamination, we removed all stars with S/N $< 45$ and log surface gravities ($\log g$) $< 4.0$.  The resulting combined catalog contains 1,148 planets around 716 stars.

\subsection{Planets and System Architectures}
We cross-matched the stellar catalog with the confirmed planet catalog from the NASA Exoplanet Archive and adopted those planet parameters.  We then defined three classes of exoplanet system architecture: hot-Jupiters, cool-Jupiters, and compact multi-planet systems.  Hot-Jupiter systems are defined as having a planet with $M_{\mathrm{planet}} > 0.5 M_{\mathrm{Jupiter}}$ or $R_{\mathrm{planet}} > 0.75 R_{\mathrm{Jupiter}}$ and semi-major axis $<=0.3$ au resulting in 104 hot-Jupiter systems. Cool-Jupiters have the same mass or radius definition as hot-Jupiter planets, but have semi-major axes $>0.3$ au.  This results in 87 cool-Jupiter systems.  Finally, we defined compact multi-planet systems as having three or more planets orbiting at less than 1~au, resulting in 105 compact multi-planet systems.  Only one hot-Jupiter system is also defined as a compact multi-Planet system, and nine cool-Jupiter systems are in the compact multi-Planet sample.

\subsection{Planet Architecture Occurrence Analysis}
To evaluate the relative occurrence rate of each system architecture as a fraction of known planet hosts at a given metallicity, we used Gaussian kernel density estimates (KDE) for the entire sample and each sub-population as a function of [Fe/H].  The optimum bandwidth was determined using Scott's rule\citep{Anonymous:1979cw}.  From the KDEs we generated probability density functions (PDF) and then divided the PDF of each architecture by that of all known hosts.  This gives us an estimate of the fraction of planetary systems of a particular architecture given the overall occurrence of discovered planetary systems as a function of metallicity.  To visualize the uncertainties in these occurrence rates, we drew 200 bootstrap realizations from each of the architecture samples and performed the same procedure for these samples.  We then calculated the 68\% and 95\% confidence regions over the entire metallicity range based on those bootstrap realizations.

\section{Results}
\begin{figure*}[htbp!] %  figure placement: here, top, bottom, or page
   \centering
   \includegraphics[width=\textwidth]{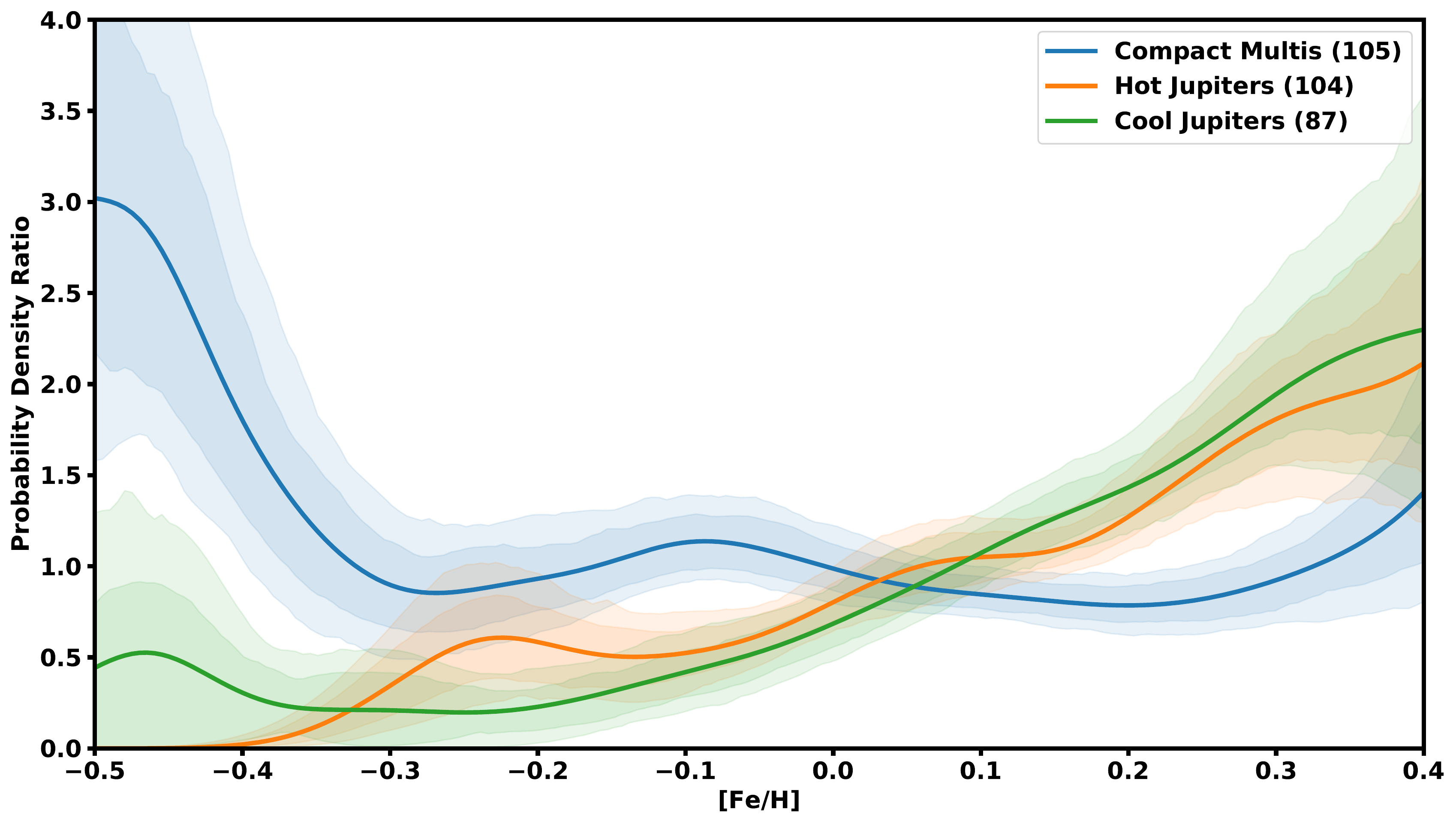}
   \caption{The frequency of compact multi-planet systems (blue) increases with decreasing metallicity as a fraction of known planet hosts.  This is in contrast to hot-Jupiter systems (orange), which are assumed to be more frequent around higher metallicity stars as a consequence of the core accretion model of planet formation.  A comparison sample of cool-Jupiters (green), systems with a giant planet on a wider orbit that may also include other planets, is more like the distribution of hot-Jupiter systems than the compact multi-planets, particularly at higher metallicities.  The bold lines represent a Gaussian kernel regression to the distribution of metallicities of the specific architecture divided by the distribution for all known planet hosts.  The shaded regions of the same color are 68\% and 95\% uncertainty intervals derived from fits to 200 bootstrap realizations of each distribution.}
   \label{fig:arch_frequency}
\end{figure*}

Comparing these planet architecture ratios, we find two opposing trends versus their log solar relative iron abundance, or [Fe/H] (Figure \ref{fig:arch_frequency}).  For hot-Jupiters, we see the expected planet-metallicity correlation.  The frequency of hot-Jupiters increases with increasing metallicity.  At the highest metallicities, the frequency of compact multi-planet systems also increases, but the frequency is also consistent with being almost flat from $-0.3 < \mathrm{[Fe/H]} < 0.3$.  However, at metallicities below -0.3 dex there is a sharp increase in the fraction of known planet hosts that are compact multi-planet systems, with a factor of three increase in the probability density over a range of just 0.2 dex.

Recent studies have suggested that cool-Jupiters, giant planets residing more than 1 AU from their host star, may be companions to compact multi-planet systems where large mutual inclinations prevent us from seeing one or the other \citep{2014ApJ...796...47M,2018ApJ...860..101Z}.  Due to their distance from their host star, it is more difficult to detect cool-Jupiters through either radial velocities or transits and as a consequence we are much less complete.  However, it is instructive to compare their distribution to the other hosts to see if they are clearly associated with one or the other population.  As a function of metallicity, cool-Jupiters seem to closely trace the behavior of the hot-Jupiter systems and metallicities higher than $\sim -0.3$.  The original planet-metallicity study included systems with planets on periods shorter than 4 years, which includes many cool-Jupiters, so this result is not too surprising.  However, for systems with $\mathrm{[Fe/H]} < -0.3$, there is an increase in cool-Jupiter frequency similar to that of the compact multi-planet systems, although less strong and driven by the nine overlapping systems between the cool-Jupiters and compact multis.

\begin{figure*}[htbp!] %  figure placement: here, top, bottom, or page
   \centering
   \includegraphics[width=\textwidth]{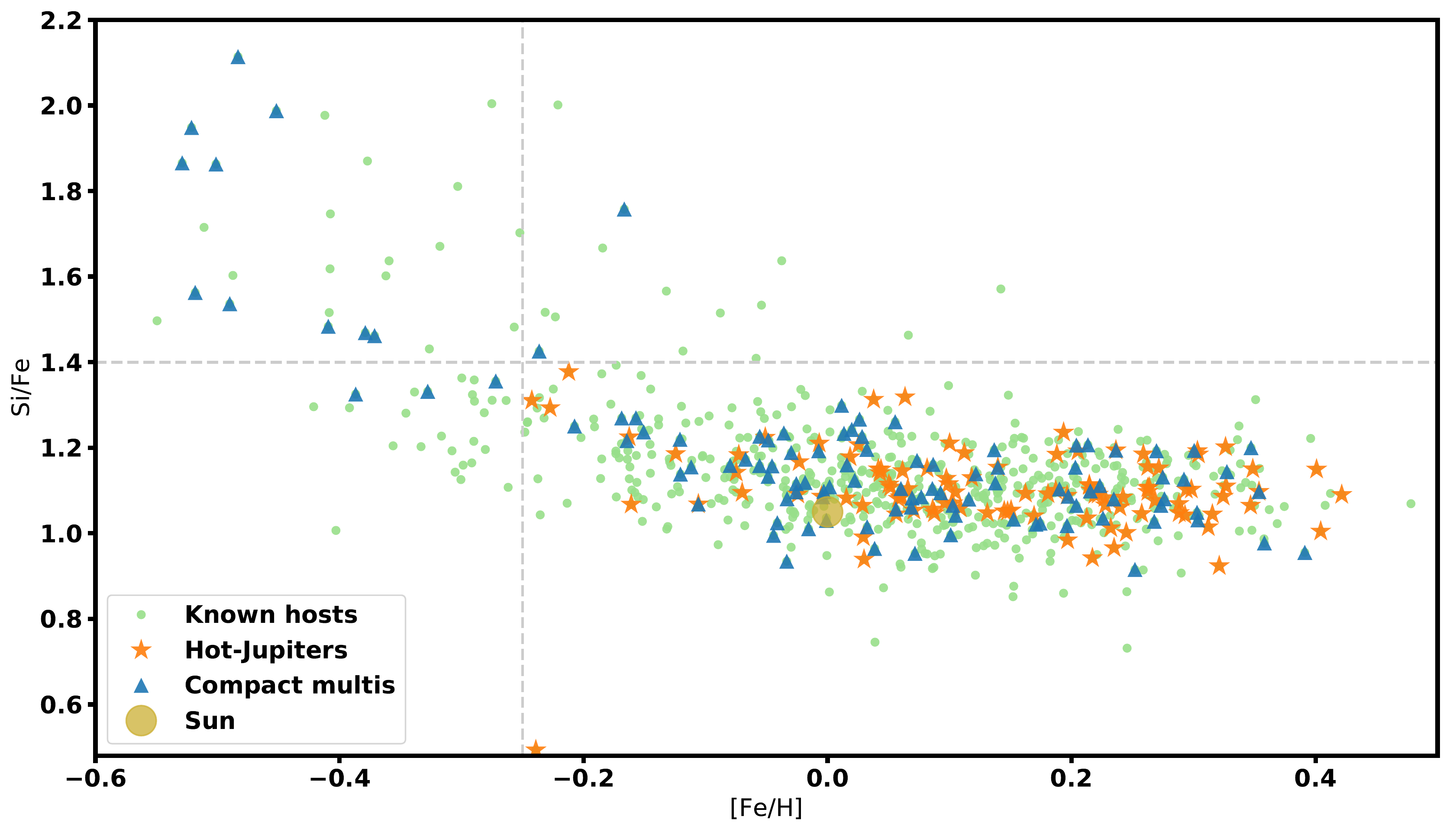}
   \caption{Stars with low metallicity or a high ratio of Si/Fe do not seem to form hot-Jupiters, and are increasingly likely to host compact multi-planet systems.  Plotted here are planet-hosting main sequence stars from our sample (green points), comparing the Si/Fe ratio to the log solar relative iron abundance, [Fe/H], with hot-Jupiters (orange stars) confined to the lower right-hand side of the plot and compact multi-planet systems (blue triangles) making up a large fraction of the upper-left region.  The Sun (yellow circle) is plotted for reference and the dashed lines are drawn to highlight the different populations.  Stars with both high Si/Fe and high [Fe/H] are thought to be from the galactic thick-disk and are poorly represented in most planet search samples due largely to their greater distance.}
   \label{fig:arch_sife_feh}
\end{figure*}

As a star ages, diffusion at the base of the convective zone can result in an apparent decrease in the amount of heavy elements at the stellar surface.  The effect is more pronounced for more massive stars, but affects all elements heavier than helium roughly equally.  Stars with lower initial metallicity should also have a higher ratio of $\alpha$-elements to iron \citep{2015A&A...582A.122K}, so we can use the Si/Fe ratio as a function of [Fe/H] to see if the increase in frequency of compact multis is due to age or inherently low metallicity (Figure \ref{fig:arch_sife_feh}).  At both low metallicity and high Si/Fe, none of the hosts are hot-Jupiters and an increasing fraction are compact multi-planet systems.  Low-metallicity stars in our sample have a higher Si/Fe, as expected for stars with initially low metallicity.  The metallicity relation we see is not related to an observational bias caused by diffusion.

\section{Discussion}

Between $-0.3 < [Fe/H] < 0.4$, the fraction of systems that are compact multis stays relatively constant despite the steep increase in the fraction of hot and cool-Jupiters. Compact multis are already known to be common around hosts of solar composition \citep{2018ApJ...860..101Z}. The increasing fraction of stars hosting compact multi-planet systems at lower metallicities points to a previously unrecognized reservoir of small-planet hosts.  This has implications for planet formation models and may suggest that the these dynamically cool systems may form after the gas disk dissipates \citep{2018MNRAS.480.2206O}.  Previous studies have found evidence for two possible populations of planets, one with low mutual inclinations and low obliquities, and a second dynamically hotter one with fewer planets \citep{2013ApJ...771...11A,2016ApJ...816...66B}.  A separate study published while this Letter was being submitted also suggests that planet multiplicity may be tied to metallicity, with multi-planet systems more common around lower-metallicity stars \citep{2018arXiv180809451Z}, supporting our result.

Stars of lower metallicity and higher Si/Fe ratios are generally older or members of the galactic thick-disk population \citep{2015A&A...582A.122K}.  This could point to a changing mix of planet architectures based on formation time and location.  In fact, one of the oldest verified and low-metallicity planet hosts, Kepler-444, is home to a compact multi-planet system \citep{Campante:2015ei}.  New high precision radial velocity surveys looking for Earth-massed planets \citep{2017arXiv171105250G,2016SPIE.9908E..6TJ} may find a much larger population of small planets around these lower metallicity stars.

\acknowledgments
D. A. Fischer and J. M. Brewer gratefully acknowledge support for this work funded under NSF 1616086. S. Wang thanks the Heising-Simons Foundation for their support.  Data presented herein were obtained at the W. M. Keck Observatory from telescope time allocated to the National Aeronautics and Space Administration through the agency's scientific partnership with the California Institute of Technology and the University of California as well as time from the Yale University TAC. The Observatory was made possible by the generous financial support of the W. M. Keck Foundation. We thank the many observers of the California Planet Search, who collected the majority of the spectra for more than a decade. This research has made use of the NASA Exoplanet Archive, which is operated by the California Institute of Technology, under contract with the National Aeronautics and Space Administration under the Exoplanet Exploration Program.

The authors wish to recognize and acknowledge the very significant cultural role and reverence that the summit of Maunakea has always had within the indigenous Hawaiian community. We are most fortunate to have the opportunity to conduct observations from this mountain.

% =============================================================================
%  Bibliography
%
%\bibliography{ms}

\end{document}